\documentclass[aps,prb,twocolumn,a4paper,citeautoscript]{revtex4-1}
\usepackage{graphicx}
\usepackage{amsfonts}
\usepackage{amssymb}
\usepackage{amsmath}
\usepackage{mathrsfs}
\usepackage{dsfont}
\usepackage{xspace} 

\newcommand\noi{\noindent}
\newcommand\be{\begin{equation}}
\newcommand\ee{\end{equation}}

\newcommand\pp{$\pi$-pulse\xspace}
\newcommand\pps{$\pi$-pulses\xspace}

\begin{document}
											      
\title{Spin coherence lifetime extension in Tm$^{3+}$:YAG through dynamical decoupling}

\author{M. F. \surname{Pascual-Winter}}
\author{R.-C. Tongning}
\author{T. Chaneli\`ere}
\author{J.-L. Le Gou\"et}
\affiliation{Laboratoire Aim\'e Cotton, CNRS-UPR 3321, Univ. Paris-Sud, B\^{a}timent 505, Campus Universitaire, 91405 Orsay cedex, France.}

\begin{abstract} 

We report on spin coherence lifetime extension on Tm$^{3+}$:YAG obtained through dynamically decoupling the thulium spins from their magnetic environment. The coherence lifetime reached with a Carr-Purcell-Meiboom-Gill sequence revealed a $450$-fold extension [$\sim$$(230 \pm 30)$ ms] with respect to previously measured values. Comparison to a simple theoretical model allowed us to estimate the correlation time of the fluctuations of the ground level transition frequency to $(172 \pm 30)$ $\mu$s at $1.7$ K. For attaining efficient decoupling sequences, we developed a strategy inspired by the \emph{zero-first-order Zeeman} effect to minimize the large inhomogeneous broadening of the ground level spin transition. 
\end{abstract}    
\maketitle


\section{Introduction}

The building block of quantum information is the superposition state between two states of a quantum system, known as a qubit. The success of a certain qubit technology depends strongly on how long the quantum character of the state holds. Reaching long coherence lifetimes of the quantum system is thus a main issue. This is even a more crucial concern when it comes to quantum memories, whose goal is to keep the information as long as possible, respecting faithfully the quantum character of the stored qubit\cite{Tittel2010, Lvovsky2009}. Rare-earth-based systems are excellent prospects of quantum memories thanks to their long coherence lifetime, $T_2$, at low temperature, both of optical and hyperfine transitions. In particular, rare-earth ion doped crystals (REIDC) present the additional advantage of providing motion-less centers without any trade-off loss of $T_2$\cite{Jacquier2005}. Therefore, rare-earth long $T_2$'s can be fully exploited in REIDC since the storage time of the device is not limited by atomic motion as it is in atomic vapors. Even though storage protocols make use of the longer-lived ground-state hyperfine (spin) superposition states, extension of $T_2$ from the $\sim$$100$-$\mu$s to the $\sim$$100$-ms scale, or even the second, is desirable in REIDC.

At low temperature, spin decoherence of some materials such as Tm${3+}$:YAG is ruled by the interaction of the rare-earth spins with the fluctuating magnetic field produced by the nuclear spins of the host matrix. Different techniques have been developed to decouple the system from its environment in the seek of extending $T_2$. The one known as {\emph{spin-locking}} consists in applying a strong radio-frequency (rf) magnetic field parallel to the spins and tuned to the spin transition \cite{Solomon1959, Hartmann1962}. However, good tuning to the spin transition is not available in inhomogeneously broadened systems, such as REIDC. Besides, when the spin coherences result from the transformation of optical ones -as in quantum storage protocols- the initial parallelism between spins and rf field cannot be ensured for all the spins in the ensemble. Therefore, spin-locking extension of $T_2$ is not compatible with quantum storage.

A successful technique recently implemented\cite{Fraval2004, Lovric2011} on REIDCs makes use of the \emph{zero-first-order Zeeman} (ZEFOZ) effect. It consists in carefully looking for a static magnetic field precisely sized and oriented such that the gradient of the ground level splitting with respect to the magnetic field coordinates vanishes. As a result, the spins are insensitive (in first order) to the environment-induced magnetic field fluctuations. This has led to $T_2$ extensions of up to $3$ orders of magnitude\cite{Lovric2011}. Nevertheless, ZEFOZ critical points are available only in REIDCs that present a zero-field hyperfine splitting \cite{Fraval2004}. 

Dynamical decoupling (DD) is a broadly used technique for the extension of $T_2$ (for a review, see Ref. \onlinecite{Yang2011}). Based on the Hahn spin echo\cite{Hahn1950}, it was first applied in the context of NMR for high precision spectroscopy \cite{
Mehring1983}. More recently, the interest for preserving the coherence in quantum information qubits has been pointed out \cite{Viola1998, Ban1998, Zanardi1999, Viola1999}. The simplest, and still highly efficient, DD scheme is the Carr-Purcell-Meiboom-Gill (CPMG) sequence\cite{Carr1954, Meiboom1958}. For an initial coherence state known at $t=0$, it consists in the application of \pps separated by intervals of duration $\tau$, the first pulse arriving at $t=\tau/2$ [see Fig. \ref{CPMG_4pulses_2ms}(a)]. The phase of the \pps alternates between $0$ and $\pi$ in order to compensate for pulse imperfections. Disregarding the alternating phase, the scheme can be alternatively seen as the repetition of an elementary sequence [dashed rectangle in Fig. \ref{CPMG_4pulses_2ms}(a)] composed by a $\tau/2$-long free evolution interval, a \pp and another $\tau/2$-long free evolution interval. If the spin level splitting does not change significantly during the duration $\tau$ of the total free evolution time in the elementary sequence, the spin coherence will be found at known states at the end of each elementary sequence, i.e. at $t=j\tau$, with $j$ the elementary sequence index. These states are either the initial state, for $j$ even, or a $\pi$-phase-shifted state, for $j$ odd. Hence, the coherence can be preserved as long as the \pps are applied. The challenge consists in increasing the \pp rate in order to compensate for the effect of the environment before this one suffers from significant reconfiguration. 

Experimental realizations of DD have yielded remarkable results, especially in molecules\cite{Morton2006}, trapped ions\cite{Biercuk2009} and nitrogen-vacancy centers in diamond\cite{Du2009}. Fraval and coworkers have successfully applied the CPMG sequence to Pr$^{3+}$:Y$_2$SiO$_5$\cite{Fraval2004}, the most thoroughly studied REIDC for quantum memory applications. A combination of the ZEFOZ and CPMG techniques yielded a coherence lifetime over $30$-s long.

Thulium-doped YAG (Tm$^{3+}$:YAG) has been identified as an attractive REIDC thanks mainly to its diode-laser-accessible optical line ($793$ nm) and to its very simple level structure under static magnetic field (two ground and two excited sub-levels) \cite{Macfarlane1993}. Its large optical inhomogeneous broadening ($20$ GHz) can also turn out useful for broadband applications such as light storage\cite{Bonarota2011}. In recent years, it has been actively investigated in the prospect of quantum storage\cite{deSeze2006, Louchet2007, Louchet2008, Lauro2009, Lauro2009Jun, Chaneliere2010, Bonarota2010, Lauro2011, Bonarota2011}. However, the degeneracy of its hyperfine sublevels due to quenching of the total angular momentum \cite{Abragam1970} does not allow the ZEFOZ technique. $T_2$ extension in Tm$^{3+}$:YAG must thus rely on DD only. Nevertheless, the simple CPMG sequence cannot be straightforwardly transposed to Tm$^{3+}$:YAG. The reason, as we will discuss in detail below, is its large spin transition inhomogeneous broadening ($\sim$$20$ times broader than in Pr$^{3+}$:Y$_2$SiO$_5$), which imposes high power constraints to the rf field. Hence, a strategy for minimizing the spin inhomogeneous linewidth is compulsory.

This article reports on $T_2$ extension through CPMG-based DD on Tm$^{3+}$:YAG. In Sec. \ref{Minimization}, we will discuss the problem related to inhomogeneous broadening and we will describe a ZEFOZ-inspired strategy for minimizing the broadening. Three experimental sections follow. In Sec. \ref{Characterization} we will characterize  the system (resulting inhomogeneous broadening, non-extended $T_2$, etc). In Sec. \ref{Preliminary} we will present results on the viability of the CPMG sequence on Tm$^{3+}$:YAG. Finally, Sec. \ref{Extension} will be dedicated to $T_2$ extension through DD. The Appendix describes a theoretical model that fits our experimental data.

\section{Minimization of the inhomogeneous broadening}\label{Minimization}

A CPMG sequence is based on the application of successive rf \pps to a spin ensemble. In order to act over the whole ensemble, the hard pulse spectrum must cover that of the spins. Thus, rather short pulses are necessary. At the same time, the condition of pulse area equal to $\pi$ must be fulfilled. Therefore, the broader the atomic spectrum, the higher the requirements on pulse intensity. 

As we said, the CPMG sequence has been successfully applied to Pr$^{3+}$:Y$_2$SiO$_5$. In that crystal, the inhomogeneous hyperfine linewidth, $\Gamma_\mathit{inh}$, is as narrow as 15-30 kHz\cite{Holliday1993,Ham1998}. 
A square \pp of Rabi angular frequency $\Omega$ has a bandwidth of the order of $\Omega$. In the case of $\Gamma_\mathit{inh}$ of the order of some tens of kHz, reasonable rf powers can produce rf-pulses able to act on a bandwidth several times broader than $\Gamma_\mathit{inh}$. Pulses need not be shorter than a few tens of microseconds. In other words, efficient \pps are readily affordable for Pr$^{3+}$:Y$_2$SiO$_5$.

The scenario is quite different for Tm$^{3+}$:YAG. As reported in previous experiments\cite{Lauro2011}, $\Gamma_\mathit{inh} \simeq 500$ kHz, which makes rf \pps much more challenging. Indeed, for the experimental conditions of Ref. \onlinecite{Lauro2011}, the 280-kHz Rabi frequency obtained with a $65$-W rf power yields a \pp bandwidth far too narrow to expect an efficient CPMG sequence. Hence, a strategy for minimizing $\Gamma_\mathit{inh}$ is to be developed for Tm$^{3+}$:YAG. 

Let us first address the reason behind this large $\Gamma_\mathit{inh}$. Splitting of hyperfine sublevels is achievable in Tm$^{3+}$:YAG through the application of an external magnetic field $\mathbf{B}$. For a given $\mathbf{B}$, the energy difference between sublevels is determined by the gyromagnetic tensor $\gamma$. It is actually the strong anisotropy of this tensor in Tm$^{3+}$:YAG ($\gamma_y \gg \gamma_x,\gamma_z$; $x$, $y$, $z$ to be defined later) that is responsible for high $\Gamma_\mathit{inh}$. Let us describe in a few words how this happens (we will go through the details later): the high anisotropy of the gyromagnetic tensor entails a strong sensitivity of the sublevel splitting angular frequency $\Delta$ to the orientation of $\mathbf{B}$. Either spatial inhomogeneities in the field or variations in the lattice orientation within the sample make that the magnetic field orientation seen by the Tm$^{3+}$ ions varies slightly from one ion site to another. Because of the high anisotropy of the gyromagnetic tensor, these slight variations entail strong ones in $\Delta$. This results in a span of values for $\Delta$ that add up to give a line broadening. The strategy we propose to minimize this broadening is to seek for a particular orientation of $\mathbf{B}$ where the sensitivity of $\Delta$ to the orientation is minimum. 

Although driven by a different motivation, this approach is quite similar to the ZEFOZ technique. In that case, the procedure consists in searching for a specific $\mathbf{B}$ whose orientation \emph{and magnitude} satisfy $\nabla_B \Delta = 0$\cite{Fraval2004, Longdell2006, Lovric2011}, where $\nabla_B$ is the gradient with respect to the magnetic field coordinates. In the problem we are interested in, of course, an increase of $T_2$ through the ZEFOZ effect would be welcome, but our main concern is still to minimize $\Gamma_\mathit{inh}$ in order to maximize the efficiency of \pps. The procedure here also searches for a minimization of the sensitivity of $\Delta$ to $\mathbf{B}$. However, full vanishing of $\nabla_B \Delta$ is possible only in systems that present a zero-field hyperfine splitting\cite{Fraval2004}. This is not the case of Tm$^{3+}$:YAG, where hyperfine sublevels are degenerate because of quenching of the total angular momentum \cite{Abragam1970}. Thus, a full ZEFOZ effect is not possible in Tm$^{3+}$:YAG. Nevertheless, \emph{partial} ZEFOZ, that is, vanishing derivatives of $\Delta$ with respect to \emph{two} of the magnetic field coordinates, is still possible, as we will see in detail below.

In the presence of a magnetic field, the Hamiltonian for Tm$^{3+}$ ions takes a simple form for sites of $D_2$ point symmetry, which is the case of Tm$^{3+}$ substituting for Y$^{3+}$ in a YAG matrix. It can be expressed as a nuclear Zeeman term\cite{Macfarlane1987}

\be
\mathcal{H}^\prime = -\hslash \gamma_n \mathbf{B}_{\mathit{eff}} \cdot \mathbf{I},
\label{HnZ}
\ee

\noi with an effective magnetic field given by

\be
\mathbf{B}_{\mathit{eff}} = (\gamma_x B_x \mathbf{\hat{x}} + \gamma_y B_y \mathbf{\hat{y}} +\gamma_z B_z \mathbf{\hat{z}})/\gamma_n. 
\label{effectivefield}
\ee

\noi The quantities $\mathbf{I}$ and $\gamma_n$ are the nuclear spin operator and gyromagnetic ratio, respectively, and $\mathbf{\hat{x}}$, $\mathbf{\hat{y}}$, $\mathbf{\hat{z}}$ are the three orthogonal twofold axis of the $D_2$ point group. $\gamma_x$, $\gamma_y$, $\gamma_z$ are defined as the eigenvalues of the rank 2 tensor

\be
\gamma_{\alpha\beta} = \gamma_n \mathds{1} + \frac{2 g_J \mu_B A_J}{\hbar} \sum_{n=1}^{2J+1} \frac{\langle 0 \lvert J_\alpha \rvert n \rangle \langle n \lvert J_\beta \rvert 0 \rangle}{E_n-E_0}, 
\ee

\noi where $\mu_B$ is Bohr's magneton, $g_J$ is the Land\'e factor for the manifold of total angular momentum $J$, $A_J$ is the hyperfine interaction parameter\cite{Macfarlane1987}, and $E_n$ is the energy of the eigenstate $|n\rangle$ of the Hamiltonian comprising the free ion and crystal field terms. In the case of $D_2$ point group symmetry, $\gamma_{\alpha\beta}$ is diagonal in the basis $\{\mathbf{\hat{x}}$, $\mathbf{\hat{y}}$, $\mathbf{\hat{z}}\}$\cite{Teplov1968}. The splitting between the eigenstates of the Hamiltonian \eqref{HnZ} is given by

\begin{gather}
\Delta(B_x,B_y,B_z) = \left(\gamma_x^2 B_x^2 + \gamma_y^2 B_y^2 + \gamma_z^2 B_z^2 \right)^{1/2}, \label{Delta} \\
\Delta(B,\theta,\phi) = \Delta_B(\theta,\phi) B,
\end{gather}

\noi where $B = (B_x^2 + B_y^2 + B_z^2)^{1/2}$, and $\theta$ and $\phi$ are spherical coordinates for the polar and azimuthal angles, respectively, with respect to some convenient Cartesian reference frame. 

\begin{figure}
\begin{center}
\includegraphics[width=0.3\textwidth,angle=0]{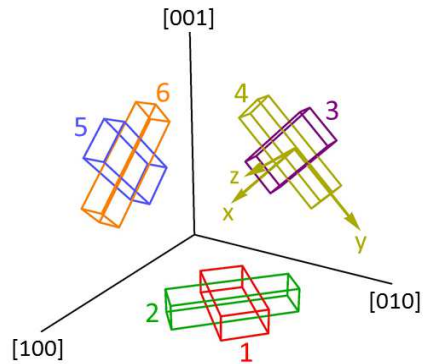}
\end{center}  
\caption{\label{Sites} (Color online). Orientations of the six orientationally distinct sites of the Y$^{3+}$ ion in the YAG crystal lattice\cite{Sun2000}. Each parallelepiped represents the local $D_2$ symmetry for one of the sites. The $\mathbf{\hat{x}}$, $\mathbf{\hat{y}}$, $\mathbf{\hat{z}}$ axis are the local axis for site 4. }
\end{figure}

\begin{figure}
\begin{center}
\includegraphics[width=0.45\textwidth,angle=0]{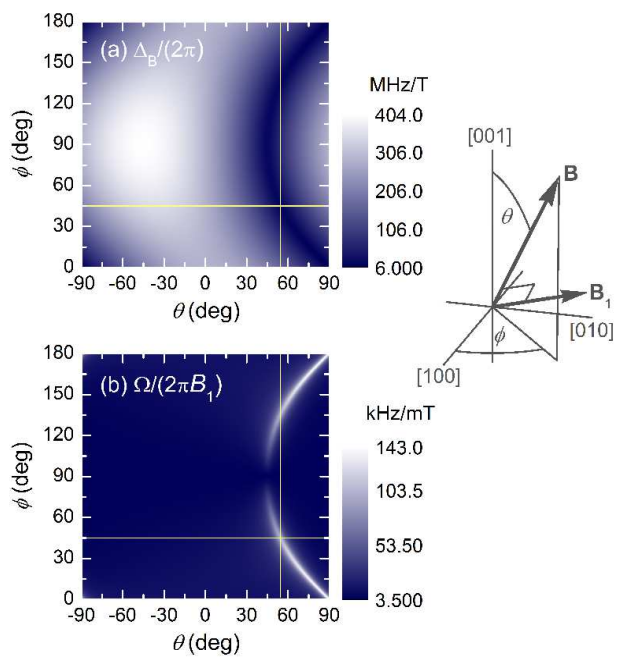}
\end{center}  
\caption{\label{SplittingRabi} (Color online). (a) Spin sublevel splitting per T and (b) Rabi frequency per mT as a function of the orientation of the static magnetic field $\mathbf{B}$ with respect to the $\{[100],[010],[001]\}$ reference frame of the YAG crystal. The rf field $\mathbf{B}_1$ is always orthogonal to $\mathbf{B}$, and contained in the $(001)$ plane. The lines indicate the orientation chosen for the experiments: $(\theta = 54.8^\circ, \phi = 45^\circ)$, which ensures low $\Gamma_\mathit{inh}$ and high $\Omega$.}
\end{figure}

Y$^{3+}$ ions in the YAG lattice occupy six sites that are equivalent in terms of their local environment but distinct in terms of the orientation of this environment, i.e. the axis of the $D_2$ point group are differently oriented (see Ref. \onlinecite{Sun2000} and references therein). The site orientations are sketched in Fig. \ref{Sites} by six parallelepiped that represent the possible orientations of the $D_2$ symmetry. When an external field is applied, its effect on each site must be treated separately. To study the behavior of $\Delta_B(\theta,\phi)$, we will consider the site for which $\mathbf{\hat{x}}$, $\mathbf{\hat{y}}$, $\mathbf{\hat{z}}$ are oriented along $[0\bar{1}\bar{1}]$, $[01\bar{1}]$, $[100]$, respectively, this is the site ``4'' as labeled in Fig. \ref{Sites}. Justification for this choice will be given later. In Fig. \ref{SplittingRabi}(a) the strong variations of  $\Delta_B$ as a function of $(\theta,\phi)$ reflect the high anisotropy of $\gamma_x$, $\gamma_y$, $\gamma_z$. For the figure, we used the values $\gamma_y = (403 \pm 3)$ MHz/T, $\gamma_x = 0.045\gamma_y$, $\gamma_z = 0.017\gamma_y$ compatible with the experimental observations of Ref. \onlinecite{deSeze2006}. The angles $\theta$ and $\phi$ have been taken with respect to the axis $\{[100],[010],[001]\}$ of the YAG Bravais lattice. 

Our aim is to minimize $\Gamma_\mathit{inh}$. For that, our strategy is to look for a magnetic field orientation $(\theta,\phi)$ for which the splitting sensitivity to the magnetic field is minimum. We see in Fig. \ref{SplittingRabi}(a) that $\Delta_B(\theta,\phi)$ displays both maxima and minima that meet this criterion. The question now is what exact field orientation to choose. For example, is it better to choose a maximum or a minimum? A minimum seems to be a better choice since inhomogeneous broadening resulting from fluctuations in the \emph{magnitude} of the magnetic field would be as small as possible. In other words, the situation would be closer to a full cancellation of $\nabla_B \Delta$. However, that is not the only criterion to take into account.

As we said before, the aim of keeping $\Gamma_\mathit{inh}$ small is to lower the constraint on \pp bandwidth, and thus on \pp duration, so that the pulses produced with the available rf power will act efficiently over the whole $\Gamma_\mathit{inh}$. Of course, rf-power is not the good parameter to look at; it is rather its associated Rabi frequency $\Omega$ which comes into play. Usually, it is the rf field component orthogonal to the static field that determines $\Omega$. For Hamiltonian terms such as \eqref{HnZ}, both for $\mathbf{B}$ and for the rf field $\mathbf{B}_1$, what counts is the component of $\mathbf{B}_{1,\mathit{eff}}$ orthogonal to $\mathbf{B}_{\mathit{eff}}$ ($\mathbf{B}_{1,\mathit{eff}}$ is defined as in Eq. \eqref{effectivefield} by substituting $B_x$, $B_y$, $B_z$ by $B_{1,x}$, $B_{1,y}$, $B_{1,z}$). In our experimental setup, the fields $\mathbf{B}$ and $\mathbf{B}_1$ are generated by two coils perpendicular to one another. As the coils are fixed, the static field orientation is actually performed by rotating the crystal rather than the magnetic field. Therefore, the orientations of $\mathbf{B}$ and $\mathbf{B}_1$ cannot be varied independently. When choosing the orientation of $\mathbf{B}$ (or rather that of the crystal) care must be taken since, because of the strong gyromagnetic tensor anisotropy, $\Omega$ may vary significantly.

In Fig. \ref{SplittingRabi}(b) we present the behavior of $\Omega$ as a function of the angles $\theta$ and $\phi$ that determine the orientation of $\mathbf{B}$ and, consequently, that of $\mathbf{B}_1$. The experimental setup fixes the rf field orientation as $\mathbf{B}_{1}/B_1 = (-\sin \phi, \cos \phi, 0)$ with respect to the crystal frame $\{[100],[010],[001]\}$. It is clear from the figure that a non-negligible $\Omega$ is observed only at very specific orientations. The region of high $\Omega$ coincides with the ``C''-shaped minimum line observed for $\Delta_B(\theta,\phi)$ in Fig. \ref{SplittingRabi}(a). This is of course not a coincidence, it results from the particular anisotropy of $\gamma_\mathit{ab}$. To understand this statement we need to bear in mind that, as soon as a given field has a non-negligible $y$ component, its associated {\emph{effective}} field is (almost) parallel to $\mathbf{\hat{y}}$ because of $\gamma_y \gg \gamma_x, \gamma_z$ [see Eq. \eqref{effectivefield}]. Thus, even though $\mathbf{B} \perp \mathbf{B}_1$, it happens that $\mathbf{B}_\mathit{eff} \parallel \mathbf{B}_{1,\mathit{eff}}$ for most field orientations. $\Omega$ is determined by the component of $\mathbf{B}_{1,\mathit{eff}}$ orthogonal to $\mathbf{B}_{\mathit{eff}}$. The orthogonality only happens if either (\emph{i}) $B_y \simeq 0$ or (\emph{ii}) $B_{1,y} \simeq 0$. In case (\emph{ii}), however, $|\mathbf{B}_{1,\mathit{eff}}|$ is strongly diminished as a result of $\gamma_x, \gamma_z \ll \gamma_y$. Therefore, only case (\emph{i}) displays high $\Omega$. At the same time, as $\gamma_y \gg \gamma_x, \gamma_z$, fulfilling (\emph{i}) results in minimizing $\Delta$ [see Eq. \eqref{Delta}]. That is why the regions of minimum $\Delta$ and high $\Omega$ coincide in Fig. \ref{SplittingRabi}.

The arguments just given answer the question on whether to choose a maximum or a minimum of $\Delta_B(\theta,\phi)$ when aiming at minimizing $\Gamma_\mathit{inh}$. To further specify the orientation $(\theta,\phi)$ of $\mathbf{B}$, it can be shown that, at $\phi = 45^\circ$, site 6 presents the same $\Delta$ and $\Omega$ as site 4 described above. This permits to double the experimental signal by addressing two sites at a time, and, at the same time, to avoid a double-valued $\Omega$. Once $\phi = 45^\circ$ set, $\theta$ is automatically fixed by selecting the minimum of $\Delta_B(\theta,45^\circ)$. It occurs at $\theta = 54.8^\circ$. For the resulting orientation $(\theta=54.8^\circ,\phi=45^\circ)$, indicated in Fig. \ref{SplittingRabi} by crossed lines, we expect $\Delta_B = 15.3$ MHz/T and $\Omega/(2\pi B_1) = 101$ kHz/mT. Static and rf fields of the order of a few T and mT, respectively, are perfectly compatible with the experimental capabilities. From previous experiments performed at $(\theta=49^\circ,\phi=45^\circ)$, we estimated the angular uncertainty in our crystal to be $\sim$$0.3^\circ$. The corresponding $\Gamma_\mathit{inh}$ obtained theoretically is $28$ kHz. The set of values we have found for $\Gamma_\mathit{inh}$, $\Delta$ and $\Omega$ make CPMG sequences experimentally feasible on Tm$^{3+}$:YAG. 

As regards the remaining sites, they will not participate in the experiment. Sites 1, 3 and 5 display a spin transition frequency far from resonance for the chosen field orientation. In addition, sites 3 and 5 present a forbidden optical dipole moment for the electric field polarization (parallel to $[11\bar{1}]$) we will use for initializing and reading out the system. That is also the case for site 2.

Here we have limited our analysis to site 4, mentioning at the end that site 6 returns equal values of $\Delta_B$ and $\Omega$ for the specific orientation $(\theta=54.8^\circ,\phi=45^\circ)$. Actually, the same situation is found for sites 3 and 5 at the orientation $(\theta=-54.8^\circ,\phi=45^\circ)$. We could have chosen to work at this orientation, just the indexes of the sites participating in the experiment would have changed. Our experimental configuration (geometry of the magnetic coils, light propagation axis and orientation of the facets of our crystal) does not allow us to work effectively with site 1 (low $\Omega$) and it does not allow any optical addressing to site 2 at all (light polarization orthogonal to the electric dipole). It is worth mentioning that, if the orientations of the magnetic fields could be varied independently and at will, and if the crystal facets could be chosen arbitrarily, any pair of sites of common index parity could be picked to participate in the experiment. Alternatively, one could choose to work with one site only, obtaining higher $\Omega$ but lower signal.

\section{Experiments and discussion}\label{Experiments}

\subsection{Characterization}\label{Characterization}

The experiments were carried out in a $0.1\%$ at Tm$^{3+}$:YAG crystal cooled down to $1.7$ K in a liquid Helium cryostat. Superconductive coils generate a static magnetic field which lifts the spin degeneracy, resulting in a four-level system composed of two ground states and two excited states. The static magnetic field is oriented as described in the previous section. As in Ref \onlinecite{Lauro2011}, the spin transition is resonantly driven by a phase-controlled rf magnetic field. A 10-turn coil, $20$-mm long and $10$-mm in diameter conveys the magnetic excitation to the crystal sitting at its center. The rf signal generated by an arbitrary wave generator (Tektronix AWG5004) passes through a pulsed amplifier (TOMCO BT00500-AlphaSA). A $\sim$$600$-kHz bandwidth rf resonant circuit further amplifies the rf current fed to the coil. The static magnetic field magnitude, close to $1$ T, is chosen to produce a ground level spitting tuned to the rf circuit resonance ($15.1$ MHz). The light beam, emerging from an extended cavity diode laser and time-shaped by acousto-optic modulators, is resonant with the optical transition of the system at $793$ nm. It is used for initializing and probing the system. It propagates along the $[\bar{1}10]$ direction with polarization parallel to $[11\bar{1}]$. The opacity of the $L = 5$-mm-long sample has been measured to be $\alpha L = 0.9$.

\begin{figure}
\begin{center}
\includegraphics[width=0.35\textwidth,angle=0]{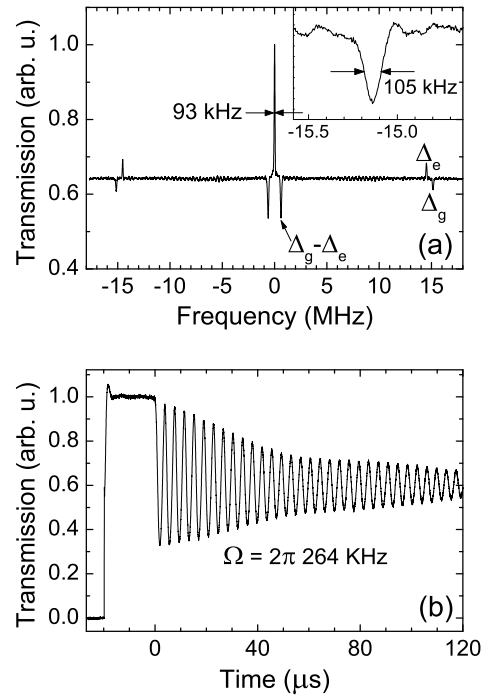}
\end{center}  
\caption{\label{HB&Nutation} (a) Hole burning spectrum. Ground level splitting antiholes ($\Delta_g$), excited level splitting holes ($\Delta_e$) and mixed antiholes ($\Delta_g-\Delta_e$) are observed in addition to the central hole. The $\Delta_g$ antihole is 105-kHz wide. (b) Spin nutation experiment. The rf field is turned on an $t=0$. The spins nutate at a Rabi angular frequency of $2\pi \times 264$ kHz.}
\end{figure}

First of all, we will experimentally verify the reduction of $\Gamma_\mathit{inh}$ and the availability of a reasonably high $\Omega$. A hole burning experiment [see Fig. \ref{HB&Nutation}(a)] revealed a full width at half maximum of the spin transition antihole of $105$ kHz. This is well below the previously reported value of $\simeq$$500$ kHz\cite{Lauro2011}, showing that the strategy for minimizing $\Gamma_\mathit{inh}$ has been fruitful. In the absence of inhomogeneous broadening, the antihole width should coincide with that of the central hole. The latter is measured to be $93$ kHz. The $\sim$$10$-kHz growth from the hole to the antihole is consistent with the expected value of $\Gamma_\mathit{inh}$. The hole width by far exceeds the homogeneous width of the optical transition, of the order of a few kHz only \cite{Macfarlane1993}. Since the hole has been deeply burnt, down to the transparency level, saturation may explain the large observed broadening, a typical feature in similar situations. The laser linewidth, reduced to less than $1$ kHz over $10$ ms by locking to a Fabry-P\'erot interferometer, should not affect the hole width, but the drift of the reference cavity during the burning process might also add to hole and antihole broadening. An imperfect orientation of $\mathbf{B}$, differing slightly from $(\theta = 54.8^\circ, \phi = 45^\circ)$, might also contribute to $\Gamma_\mathit{inh}$ larger than expected. In any case, our experimental result fixes an upper bound for $\Gamma_\mathit{inh}$. As regards $\Omega$, we have performed a spin nutation experiment [see Fig. \ref{HB&Nutation}(b)] that yielded $\Omega/(2\pi)=264$ kHz for a rf power ($\sim$$27$ W) well below the maximum capability of our amplifier. With this value, square-shaped \pps as short as $1.8$ $\mu$s are possible. This gives a pulse bandwidth, $\sim$$\Omega$, comfortably larger than $\Gamma_\mathit{inh}$. These are the parameters to be used in the sequences of the CPMG experiments to be discussed.

\begin{figure}
\begin{center}
\includegraphics[width=0.3\textwidth,angle=0]{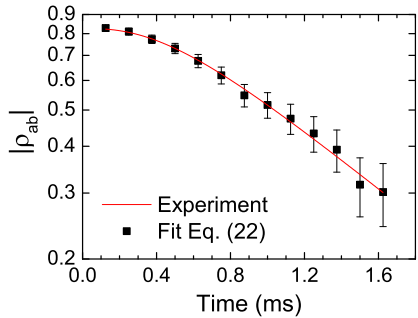}
\end{center}  
\caption{\label{T2SpinEcho} Spin echo experiment. $|\rho_\mathit{ab}|$ is obtained from the quantity $\ln(I_f/I_\mathit{ref})/\ln(I_i/I_\mathit{ref})$, where $I_i$ and $I_f$ are the transmitted probe intensities before and after the rf sequence, respectively, and $I_\mathit{ref}$ is a reference intensity for a known population difference (typically $0$). The fit returns the parameters $\sigma_\Delta/(2\pi) = (2.3 \pm 0.2)$ kHz and $\tau_c = (172 \pm 30)$ $\mu$s.}
\end{figure}

In order to weigh the impact of the CPMG sequence on the extension of $T_2$, we have measured the latter by means of an optically detected spin echo experiment. For the optical detection, a $\pi/2$-pulse is applied at the end of the spin echo sequence (see Fig. \ref{CPMG_4pulses_2ms}(a) for a sketch; consider $n=1$ ). It is aimed at converting the coherence $\rho_\mathit{ab}$ back into an optically detectable population difference. The transmission of a probe beam is measured prior and post application of the rf spin echo sequence. The relevant experimental information is contained in the comparison between the initial and final probe intensity. The results are presented in Fig. \ref{T2SpinEcho}. We observe a non-linear behavior, the non-linear character being more pronounced at short times. This tells us that, at the time scale of Fig. \ref{T2SpinEcho}, we have not yet reached the limit $t \gg \tau_c$ in which $|\rho_\mathit{ab}|$ decays in an exponential fashion ($\tau_c$ the correlation time of the fluctuations in $\Delta$ induced by the environmental aluminium spins). We also note that $|\rho_\mathit{ab}|$ does not reach unity for times approaching to zero. This can be due to imperfect matching of the rf frequency to the spin transition. Finite $\Gamma_\mathit{inh}$ also contributes to matching imperfections. We have developed a theoretical model for calculating the evolution of $\rho_\mathit{ab}$ during the CPMG sequence. Details can be found in the Appendix. This model reduces to the case of the spin echo when the number of \pps in the sequence is set to one. The expression for $\rho_\mathit{ab}$ is given in Eq. \eqref{gammaSE}. The parameters of the model are $\tau_c$ and the standard deviation of the fluctuation in $\Delta$, $\sigma_\Delta$ (a global offset is also included as a parameter). We have fitted the model to the experimental data. The result is represented by the line in Fig. \ref{T2SpinEcho}. The agreement with the experiment is satisfactory. The fitting procedure yields $\sigma_\Delta/(2\pi) = (2.3 \pm 0.2)$ kHz and $\tau_c = (172 \pm 30)$ $\mu$s. Not only are these values interesting themselves, also they permit to estimate $T_2$ even if the time scale of our experiment is not long enough to appreciate the exponential decay. The estimation is performed through the expression $T_2^{-1} = \sigma_\Delta^2 \tau_c$, known for stationary Markovian gaussian processes, and also deduced in the Appendix [Eq. \eqref{gammaSElimit}]. We get $T_2 = 1.01$ ms. This value doubles the one previously obtained for the magnetic field orientation $(\theta = 45^\circ, \phi = 49^\circ)$\cite{Lauro2011}. The rise is due to the partial ZEFOZ effect discussed in the previous section. It is worth noting that $\sigma_\Delta$ is a parameter dependant on the magnetic field orientation. On the other hand, $\tau_c$ is independent, providing us with a more general piece of information about the system.

With the information gathered so far, we know that \pps spectrally wide enough and much shorter than $\tau_c$ are feasible in Tm$^{3+}$:YAG. This enables the design of CPMG sequences aimed at extending $T_2$. In the next section, we will show initial evidence that this is indeed possible.

\subsection{CPMG experiments with a few $\pi$-pulses}\label{Preliminary}

\begin{figure}
 \begin{center}
 \includegraphics[width=0.45\textwidth,angle=0]{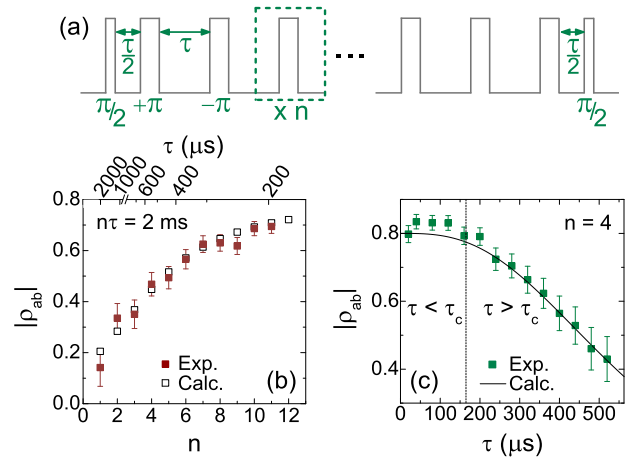}
 \end{center}  
 \caption{\label{CPMG_4pulses_2ms} (a) Scheme of the CPMG sequence. The signes ``$+$'' and ``$-$'' that precede the label of the \pps refer to the pulse phase, corresponding to $0$ or $\pi$, respectively. $n$ and $\tau$ stand for the number of \pps and the time interval between them. (b, c) Coherence recovered after a CPMG sequence with (b) fixed free evolution time $n \tau = 2$ ms and increasing $n$, or (c) fixed $n=4$ and increasing $\tau$. The calculated data are obtained from Eq. \eqref{gamma3} with $\sigma_\Delta/(2\pi) = 2.3$ kHz and $\tau_c = 172$ $\mu$s.}
 \end{figure}

 First we prepare the system by optically pumping the crystal into a single energy sublevel of the nuclear spin over an optical interval $\sim$$400$-kHz broad. Then, we apply a CPMG sequence as sketched in Fig. \ref{CPMG_4pulses_2ms}(a). It starts by a $\pi/2$-pulse which transforms the initial population difference into a coherence (rotation of the Bloch vectors from the vertical axis to the transverse plane of the Bloch sphere). A sequence of $n$ \pps follows, arriving at instants $t_j = (2j+1) \tau/2$ ($j = 0, \ldots, n-1$) with phases $[1+(-1)^{j+1}] \pi/2$. They rephase the spins at instants $j\tau$, $\tau$ being the time interval between two pulses. Just as in the spin echo experiment, another $\pi/2$-pulse of phase $[1+(-1)^{n}] \pi/2$ is applied at the end of the sequence for enabling the optical detection of the final coherence. Here again, $|\rho_\mathit{ab}|$ is recovered from the initial (right after optical pumping) and final transmitted probe intensity.
 
We have performed two preliminary experiments to test the action of the CPMG sequence onto $\rho_\mathit{ab}$. In the first one, $n$ was increased while the total free evolution time of the sequence, $n \tau$, was kept fixed at $2$ ms. The results are exhibited in Fig. \ref{CPMG_4pulses_2ms}(b). We observe that $|\rho_\mathit{ab}|$ rises as $n$ is increased ($\tau$ is shortened), providing evidence of the preservation of $\rho_\mathit{ab}$ by the CPMG sequence. Our theoretical model (hollow squares) matches the experimental data (solid squares) very well for the same $\tau_c$ and $\sigma_\Delta$ obtained from the fit of the spin echo data (Fig. \ref{T2SpinEcho}). For $n$ large, $|\rho_\mathit{ab}|$ seems to tend to a stationary behavior. In fact, the rightmost data point in Fig. \ref{CPMG_4pulses_2ms}(b) is close to $\tau_c$. A plateau-like behavior is better defined in Fig \ref{CPMG_4pulses_2ms}(c). Here, $n$ is fixed to $4$ and $\tau$ is varied through $\tau_c$. We see that, as $\tau$ is shortened, $|\rho_\mathit{ab}|$ increases until it reaches a plateau ($|\rho_\mathit{ab}|$ not attaining unity for $\tau \rightarrow 0$ is probably due to the reasons mentioned in the analysis of Fig. \ref{T2SpinEcho}). This behavior allows us to distinguish between two regimes. In fact, the borderline is in good agreement with the previously obtained value $\tau_c = 172$ $\mu$s. For $\tau > \tau_c$, $\Delta$ fluctuates significantly during one elementary sequence of the CPMG process [dashed rectangle in Fig. \ref{CPMG_4pulses_2ms}(a)]. In such a case, the CPMG strategy fails at keeping $\rho_\mathit{ab}$. For $\tau < \tau_c$, the elementary sequence is short enough to ensure that the decoherence is suppressed or, at least, strongly diminished.

\subsection{Extension of the coherence lifetime}\label{Extension}

\begin{figure}[h!]
\begin{center}
\includegraphics[width=0.35\textwidth,angle=0]{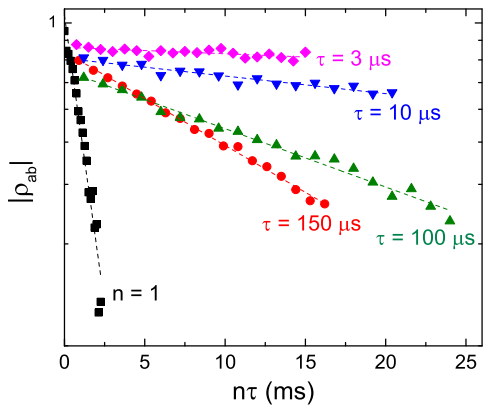}
\end{center}  
\caption{\label{Slopes} (Color online). Measurement of $T_2$ through CPMG sequences with different $\tau$ values. For data sets labeled $\tau = 150$, $100$, $10$, $3$ $\mu$s, $\tau$ is kept fixed and $n$ is increased. The data set labeled $n=1$ is a standard spin echo experiment ($n$ is fixed and $\tau$ is increased). The dashed lines result from least squares fits.}
\end{figure}

Let us now turn to experiments aimed at keeping the coherence over long time intervals. In Fig. \ref{Slopes} we show $\rho_\mathit{ab}$ as a function of $n\tau$ (total free evolution time), for CPMG sequences with a large number of pulses, and for different values of $\tau$. Each data set labeled $\tau = 150$, $100$, $10$, $3$ $\mu$s has been obtained by increasing $n$ for a fixed value of $\tau$. In the figure we also include the standard spin echo measurement (labeled $n=1$) for comparison. We observe that the data sets display a linear behavior in log scale. This translates the fact that the time scale involved is much longer than $\tau_c$ [see the Appendix, especially Eq. \eqref{calcT2}]. The slope, defined as $T_2^{-1}$, decreases significantly for decreasing $\tau$. The maximum $T_2$ attained is $(230 \pm 30)$ ms, for $\tau=3$ $\mu$s. That is over $220$ times larger than the value obtained through the standard spin echo, and $450$ times larger than the one measured for a magnetic field orientation $(\theta = 49^\circ, \phi=45^\circ)$ that does not satisfy any $\partial_\alpha \Delta=0$ ($\alpha = \theta, \phi$) condition\cite{Lauro2011} (absence of partial ZEFOZ effect). These results prove the $T_2$ extension in Tm$^{3+}$:YAG provided by the application of a CPMG sequence with $\tau < \tau_c$. 

\begin{figure}[h!]
\begin{center}
\includegraphics[width=0.4\textwidth,angle=0]{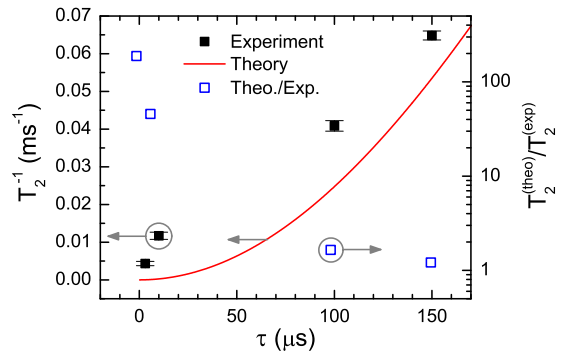}
\end{center}  
\caption{\label{T2s} (Color online). CPMG-extended $T_2$ as a function of $\tau$. The experimental values (solid black symbols) result from the least squares fit of the data of Fig. \ref{Slopes}. The theoretical prediction (solid red line) represents Eq. \eqref{calcT2}. The green dashed line indicates the lower bound $(2T_1)^{-1}$ ($T_1\sim 1$ min). The ratio between theoretical and experimental $T_2$ values is also plotted (hollow blue symbols; these data refer to the right vertical axis).}
\end{figure}

In Fig. \ref{T2s}, we plot the values of $T_2$ obtained from a least squares fit to the data of Fig. \ref{Slopes}. The figure also displays the theoretical prediction for $T_2$ according to Eq. \eqref{calcT2}. We see that the overall experimental and theoretical behaviors are similar. However, the experimental $T_2$ is always lower than the predicted one. For $\mu=150$ $\mu$s, experimental and theoretical values are $15.4$ and $18.7$ $m$s, respectively. However, for $\mu=3$ $\mu$s, the corresponding values are $0.230$ and $43$ s. The ratio theory/experiment is not far from unity for large values of $\tau$, but it attains $190$ for $\tau = 3$ $\mu$s. This is partly due to the cumulated imperfections of the spin rephasing by one single elementary sequence. For $\tau = 150$ $\mu$s, only $100$ elementary sequences are applied to reach $15$ ms of free evolution, but for $\tau = 3$ $\mu$s, $5000$ sequences are necessary. In addition to this source of discrepancy between experiment and theory, it is worth recalling that we have made use of a very simple model of decoherence in virtue of its straightforward mathematical treatment (see the Appendix). This model assigns decoherence to fluctuations of the spin transition frequency only. The latter quantity is defined as a stationary, Markovian, gaussian stochastic process, three properties that suffice for complete characterization. We estimate that spin-spin interactions between the impurity ion and the spin bath of the crystal (other than those of the \emph{flip-flop} and frozen core kinds) can be reasonably classed as stationary and Markovian. However, the qualification of gaussian, taken only because of the analytical resolution it provides, is questionable and would need further consideration. In addition, spin-lattice interactions, for instance, of rarer occurrence and responsible for $T_1$ (measured to be $\sim 1$ min), are not at all taken into account in our model. Flip-flop and frozen core interactions are discarded as well. When looking at long times, which becomes possible thanks to the DD compensation for spin-spin decoherence, these effects become dominant. This, together with the uncertainty of the gaussian assumption, might explain the discrepancy with the experimental results. Other relaxation sources not accounted for in our calculations can be invoked such as crystal heating or instantaneous spectral diffusion by the rf pulses. Further investigation is needed to estimate the potential impact of those effects.

\section{Conclusions}

$\pi$-Pulse-based DD techniques on Tm$^{3+}$:YAG posed the problem of high rf power requirements imposed by the large inhomogeneous broadening of the spin transition. We have shown that it is possible to minimize this broadening by carefully choosing an orientation for the static magnetic field such that the sensitivity of the spin transition frequency to the field orientation is minimal. This way, a large span in magnetic field orientation seen by spins at different sites in the lattice translate into a small span of transition frequencies. The inhomogeneous broadening has been reduced by a factor $\geq 5$. At the same time, $T_2$ has been doubled as a consequence of a partial ZEFOZ effect. $\pi$-Pulses effective over the whole inhomogeneous line are now feasible on Tm$^{3+}$:YAG with reasonable rf powers. This has enabled the application of a CPMG DD sequence. We have observed a clear increase in $T_2$ as the time length of the elementary sequence in the CPMG protocol was reduced. A maximum value of $(227 \pm 30)$ ms was obtained. It represents a $445$-fold rise with respect to previous experimental values. A simple theoretical model we proposed showed excellent agreement to our experimental data for sequences with few pulses ($\sim$$10$). It allowed the extraction of the parameters that characterize the fluctuation: its standard deviation and its correlation time, estimated to $(2.3 \pm 0.2)$ kHz and $(172 \pm 30)$ $\mu$s (at $1.7$ K), respectively. The comparison between experiment and model declines as the number of pulses is increased. We assume this is due to accumulation of pulse imperfections and to decoherence mechanisms disregarded in the model. 

\section{Acknowledgements}

This research has been supported by the European Commission through FP7-QuReP(STREP-247743) and FP7-CIPRIS(MC ITN-287252),  by the Agence Nationale de la Recherche through ANR-09-BLAN-0333-03 and by the Direction G\'en\'erale de l'Armement.

\section*{Appendix}

Here we aim at calculating the evolution of the coherence during a CPMG sequence. We will consider an inhomogeneously broadened atomic transition centered at $\Delta_0$. Each ion in the inhomogeneous line suffers from crystal field fluctuations that induce fluctuations in its spin transition frequency. Its detuning from the line center varies as

\be
\delta (t) = \bar{\delta} + \zeta(t),
\ee 

\noi where $\bar{\delta}$ is the time-averaged detuning. We assume the initial situation ($t=0$) in which the coherence $\rho_\mathit{ab}(t=0) = \rho_\mathit{ab}^{(0)}$ is the same for all the ions, and there is no population difference (all Bloch vectors aligned along an axis contained in the transverse plane of the Bloch sphere). This scenario can be reached, for example, by emptying one of the levels of the spin transition (optical pumping) and applying a rf $\pi/2$-pulse, as done in the experiments described above.  

The CPMG sequence consists of a succession of $n$ \pps. The \pps are applied at instants $t=(2j+1)\tau/2$, $j=0,\ldots,n-1$. We are interested in the echo intensity at $t= n \tau$. This is proportional to the coherence averaged over the inhomogeneous distribution, $\tilde{\rho}_{ab}$. As each \pp of the sequence conjugates the coherences, in the reference frame rotating at angular frequency $\Delta_0$ $\tilde{\rho}_{ab}$ is given by

\begin{widetext}
\begin{eqnarray}
\tilde{\rho}_{ab} (n, \tau) &=& \rho_\mathit{ab}^{(0)}\int d\bar{\delta} g(\bar{\delta}) \Bigg\langle \exp\Bigg[i \int_{(2n-1)\frac{\tau}{2}}^{n\tau} dt \delta(t) -\bigg(i \int_{(2n-3)\frac{\tau}{2}}^{(2n-1)\frac{\tau}{2}} dt \delta(t) -\bigg\{\dots - \bigg[i \int_{3\frac{\tau}{2}}^{5\frac{\tau}{2}} dt \delta(t) -\bigg(i \int_{\frac{\tau}{2}}^{3\frac{\tau}{2}} dt \delta(t)\nonumber\\
&& -i \int_0^{\frac{\tau}{2}} dt \delta(t) \bigg)\bigg]\bigg\}\bigg)\Bigg] \Bigg\rangle,
\label{echo1}
\end{eqnarray}

\begin{eqnarray}
\tilde{\rho}_{ab} (n, \tau) &=& \rho_\mathit{ab}^{(0)} \Bigg\langle \exp\Bigg\{i \Bigg[\int_{(2n-1)\frac{\tau}{2}}^{n\tau} dt \zeta(t)  +(-1)  \int_{(2n-3)\frac{\tau}{2}}^{(2n-1)\frac{\tau}{2}} dt \zeta(t)+\dots + (-1) ^{n-2} \int_{3\frac{\tau}{2}}^{5\frac{\tau}{2}} dt \zeta(t) \nonumber\\
&& +(-1) ^{n-1} \int_{\frac{\tau}{2}}^{3\frac{\tau}{2}} dt \zeta(t) +(-1)^n \int_0^{\frac{\tau}{2}} dt \zeta(t) \Bigg] \Bigg\} \Bigg\rangle,
\label{echo2}
\end{eqnarray}
\end{widetext}

\noi where $g(\bar{\delta})$ stands for the inhomogeneous distribution and $\langle\ldots\rangle$ represents the statistical average. For going from Eq. \eqref{echo1} to Eq. \eqref{echo2} we have taken into account that the contribution from $\bar{\delta}$ cancels out. Equation \eqref{echo2} can be written as

\be
\tilde{\rho}_{ab} (n, \tau) = \rho_\mathit{ab}^{(0)} \Big\langle e^{i \int_0^{n \tau} dt\ s_\tau(t) \zeta(t)} \Big\rangle,
\ee

\noi with

\be
s_\tau(t) = \left\{ \begin{array}{cl}
(-1)^n & 0 < t < \frac{\tau}{2}, \\
(-1)^{n-j} & (2j-1)\frac{\tau}{2} < t < (2j+1)\frac{\tau}{2} \\
1 & (2n-1)\frac{\tau}{2} < t < n\tau
\end{array} \right. ,
\label{st}
\ee

\noi where $j=1,\ldots, n-1$.

Let $\zeta(t)$ be a stationary, gaussian, Markovian process. According to Doob's theorem \cite{Doob1942}, only the Ornstein-Uhlenbeck \cite{Uhlenbeck1930} process simultaneously satisfies those three assumptions. The corresponding autocorrelation function reads

\be
\langle \zeta(t) \zeta(t^\prime) \rangle = \sigma_\Delta^2 e^{-|t-t^\prime|/\tau_c},
\ee

\noi where $\sigma_\Delta$ and $\tau_c$ represent the standard deviation of the fluctuation and the correlation time, respectively. Since $\zeta(t)$ is a gaussian process, it can be shown that

\be
\big\langle e^{i \int_{t_0}^{t} dt^\prime f(t^\prime) \zeta(t^\prime)} \big\rangle = e^{-\frac{1}{2} \big\langle \left[  \int_{t_0}^{t} dt^\prime f(t^\prime) \zeta(t^\prime) \right]^2 \big\rangle}.
\ee

\noi With this in mind, the expression for the coherence reduces to

\be
\tilde{\rho}_{ab} (n, \tau) = \rho_\mathit{ab}^{(0)} e^{-\gamma(n, \tau)},
\ee

\noi with

\be
\gamma(n, \tau) = \frac{1}{2} \Bigg\langle \left[  \int_{0}^{n \tau} dt^\prime s_\tau(t) \zeta(t) \right]^2 \Bigg\rangle.
\ee

Let us perform the calculation for $\gamma(n, \tau)$.

\begin{eqnarray}
\gamma(n, \tau) &=& \frac{1}{2} \int_0^{n\tau} dt  \int_0^{n\tau} dt^\prime s_\tau(t) s_\tau(t^\prime) \langle \zeta(t) \zeta(t^\prime) \rangle, \nonumber\\
&=& \frac{\sigma_\Delta^2}{2} \int_0^{n\tau} dt  \int_0^{n\tau} dt^\prime s_\tau(t) s_\tau(t^\prime) e^{-|t-t^\prime|/\tau_c}, \\
\end{eqnarray}

\noi The above double integral simplifies if we consider the relation $\int_{t_0}^t dt^\prime \int_{t_0}^t dt^{\prime\prime} f(t^\prime,t^{\prime\prime}) = 2 \int_{t_0}^t dt^\prime \int_{t_0}^{t^\prime} dt^{\prime\prime} f(t^\prime,t^{\prime\prime})$, valid for any $ f(t^\prime,t^{\prime\prime})$ symmetric after permutation of $t^{\prime}$ and $t^{\prime\prime}$. We therefore get

\begin{eqnarray}
\gamma(n, \tau)  &=&  \sigma_\Delta^2 \int_0^{n\tau} dt  \int_0^{t} dt^\prime s_\tau(t) s_\tau(t^\prime) e^{-(t-t^\prime)/\tau_c}, \\
&=& \sigma_\Delta^2 \int_0^{n\tau} dt\ s_\tau(t) h_\tau(t), \label{gamma1}
\end{eqnarray}

\noi with

\begin{widetext}
\be
h_\tau(t) = (-1)^n (\sigma_\Delta\tau_c)^2 e^{-\frac{t}{\tau_c}}
\begin{cases}
\begin{split}
(e^{\frac{t}{\tau_c}}-1) 
\end{split} & 0 < t < \frac{\tau}{2} \\ 
\begin{split}
\bigg\{ e^{\frac{\tau}{2\tau_c}} -1 -\frac{e^{\frac{\tau}{\tau_c}}-1}{e^{\frac{\tau}{\tau_c}}+1} e^{\frac{\tau}{2\tau_c}}  \left[ 1+(-1)^{j} e^{(j-1)\frac{\tau}{\tau_c}} \right] \\ 
{} + (-1)^{j} \left( e^{\frac{t}{\tau_c}}-e^{(2j-1)\frac{\tau}{2\tau_c}} \right) \bigg\}  
\end{split} 
& (2j-1)\frac{\tau}{2} < t < (2j+1)\frac{\tau}{2} \\
\begin{split}
\bigg\{ e^{\frac{\tau}{2\tau_c}} -1 -\frac{e^{\frac{\tau}{\tau_c}}-1}{e^{\frac{\tau}{\tau_c}}+1} e^{\frac{\tau}{2\tau_c}}  \left[ 1+(-1)^{n} e^{(n-1)\frac{\tau}{\tau_c}} \right] \\ 
{} + (-1)^{n} \left( e^{\frac{t}{\tau_c}}-e^{(2j-1)\frac{\tau}{2\tau_c}} \right) \bigg\}  
\end{split} 
& (2n-1)\frac{\tau}{2} < t < n\tau \\
\end{cases}
,
\label{h}
\ee
\end{widetext}

\noi where $j$ runs from $1$ through $n-1$. By computing the integral in Eq. \eqref{gamma1}, we obtain the final expression for $\gamma(n, \tau)$:

\be
\begin{split}
\gamma(n, \tau) = &\left(\sigma_\Delta\tau_c\right)^2 \Bigg\{ \left[ \frac{1}{\tau_c} - \frac{2}{\tau} \tanh\left(\frac{\tau}{2\tau_c}\right) \right] t \\
& - \Big[ 1 + (-1)^{n+1} e^{-t/\tau_c}\Big] \left[ 1 - \mathrm{sech}\left(\frac{\tau}{2\tau_c}\right) \right]^2 \Bigg\}.
\end{split}
\label{gamma3}
\ee

\noi It is understood in Eq. \eqref{gamma3} that $t=n\tau$.

We see from Eq. \eqref{gamma3} that the coherence does not decay in an exponential manner. However, the exponential behavior is recovered in the $t \gg \tau_c$ limit with a characteristic time given by

\be
T_2^{-1}(\tau) = \sigma_\Delta^2 \tau_c \left[ 1 - \frac{2\tau_c}{\tau} \tanh\left(\frac{\tau}{2\tau_c}\right) \right]
\quad,\quad t \gg \tau_c.
\label{calcT2} 
\ee

\noi Moreover, if $\tau \ll \tau_c$, $T_2$ reduces to

\be
T_2^{-1}(\tau) = \frac{1}{12} \frac{\sigma_\Delta^2 \tau^2}{\tau_c}
\quad,\quad t \gg \tau_c \gg \tau.
\ee

From Eq. \eqref{gamma3}, we can easily calculate the coherence lifetime obtained by a single spin echo by assigning $n=1$ and $\tau = t$. This yields

\be
\gamma_{se}(t) = \left(\sigma_\Delta\tau_c\right)^2 \left( \frac{t}{\tau_c} + 4 e^{-\frac{t}{2\tau_c}} - e^{-\frac{t}{\tau_c}} - 3 \right) .
\label{gammaSE}
\ee

\noi In the limit $t \gg \tau_c$ we get the standard result

\be
\gamma_{se}(t) = \sigma_\Delta^2 \tau_c t \quad , \quad t \gg \tau_c.
\label{gammaSElimit}
\ee

It is worth mentioning that the development described here to obtain $\gamma(n,\tau)$ is analog to the extension to $n$ \pps of the calculation performed by Herzog and Hahn in Ref. \onlinecite{Herzog1956} for one \pp. The same result is found. 


%

\end{document}